\documentclass[%
 reprint,
 amsmath,amssymb,
 aps,
]{revtex4-1}

\usepackage{graphicx}
\usepackage{dcolumn}
\usepackage{bm}
\usepackage{hyperref}

\usepackage{xcolor}

\newcommand{\re}{\text{Re}}
\newcommand{\im}{\text{Im}}

\begin{document}


\title{Zeros of partition functions in the NPT-ensemble}
\author{Timur Aslyamov}
\email{t.aslyamov@skoltech.ru}

\author{Iskander Akhatov}
\affiliation{Center for Design, Manufacturing and Materials,
   Skolkovo Institute of Science and Technology,
  Bolshoy Boulevard 30, bld. 1, Moscow, Russia 121205}

\date{\today}

\begin{abstract}
Lee-Yang and Fisher zeros are crucial for the study of phase transitions in the grand canonical and the canonical ensembles, respectively. However, these powerful methods do not cover the isothermal-isobaric ensemble (NPT ensemble), which reflects the conditions of many experiments. In this work we present a theory of the phase transitions in terms of the zeros of the NPT-ensemble partition functions in the complex plane. The proposed theory provides an approach to calculate all the partition function zeros in the NPT ensemble, which form certain curves in the thermodynamic limit. To verify the theory we consider Tonks gas and van der Waals fluid in the NPT ensemble. In the case of Tonks gas, similarly to the Lee-Yang circle theorem, we obtain an exact equation for the zero limit curve. We also derive an approximated limit curve equation for van der Waals fluid in terms of the Szegö curve. This curve fits numerically calculated zeros and correctly describes how the phenomenon of phase transition depends on the temperature.   
\end{abstract}

\maketitle

\section{Introduction}
In two seminal works \cite{lee1952statistical,yang1952statistical} Lee and Yang demonstrated that a phase transition in the open systems is described by the distribution of the partition function zeros in the complex fugacity plane. More precisely, Lee and Yang shown that in the thermodynamic limit the zeros form a circle and tend to a point on positive fugacity-axis inducing phase transition. Therefore, an information about the the partition function zeros distribution of finite number of molecules is crucial to calculate singularities at thermodynamic limit. Considering the finite (small) systems, the thermodynamic properties strongly depend on used statistical ensemble \cite{hill1994thermodynamics, dunkel2006phase, aslyamov2014complex, aslyamov2014some}. Hence, constant number of molecules at canonical ensemble is described by Fisher zeros in the plane of complex temperature \cite{fisher1965statistical}, which exhibit more complicated limit behaviour than Lee-Yang zeros \cite{bena2005statistical}. Lee-Yang and Fisher zeros are powerful approach to investigate a various list of many body systems in grand canonical and canonical ensembles, respectively \cite{bena2005statistical}. 

Another important class of the systems correspond to NPT ensemble (isothermal-isobaric ensemble), where $N$ the number of molecules, $P$ the pressure and $T$ the temperature are constant, that closely reflects the conditions of many condensation experiments \cite{tuckerman2010statistical}. However, the literature review shows that NPT-ensemble does not have description in terms of partition function zeros. Also strong motivation of our interest to NPT-ensemble is related to its wide distribution, especially at nanoscale. One of the examples is graphene nanobubbles (van der Waals heterostructures) which play an important role in the new materials science \cite{geim2013van}. Actually experimentally observed graphene nanobubbles \cite{khestanova2016universal} store trapped fluid molecules and variate the confinement profile (the volume) due to graphene elasticity \cite{iakovlev2017atomistic}. Therefore a graphene nanobubble at NPT conditions takes certain form and volume to reach the equilibrium, which significantly depends on stored fluid \cite{zhilyaev2019liquid, iakovlev2019modeling}. Accounting for the small number of trapped molecules proposed here the NPT ensemble rigorous description is needed. 

In this work we propose a new theory to provide information about the zeros of NPT ensemble partition function in the complex plane. We considered the interacting molecules using Mayer cluster expansion approach that leads to virial equation of state for real fluids \cite{hill1987statistical}. The connection of the divergence of activity expansion and phase transition phenomenon is the object of recent studies \cite{ushcats2017divergence,ushcats2018evidence}
Our developed theory of phase transitions in the NPT-ensemble avoids the low-density limitation of the original Mayer's approach due to convergence radius of the series for density and activity. Previously statistical analysis of Mayer expansion was extended only in the canonical ensemble \cite{ushcats2012equation}. We demonstrated that the phase transitions in NPT ensemble are induced by behaviour of the curve fitting the partition function zeros in the thermodynamic limit. To verify proposed theory we considered one dimensional hard sphere Tonk's gas and the real fluids van der Waals (vdW) EoS. In the case of Tonk's gas the obtained limit curve coincides with exact solution in terms of Szego curve. For vdW fluid we developed an accurate approximation of the limit curve which correctly reflects temperature dependence, recognizing the difference between critical and subcritical conditions. 
\section{Method}

NPT-ensemble was introduced by Guggenheim as summation of a partition function in canonical ensemble over discrete volume  \cite{guggenheim1939grand}. Now the NPT-ensemble is used in continuum limit and can be written as the following integral \cite{hill1987statistical}:
\begin{eqnarray}
\label{eq:PF_IIE_1}
\Delta_N(\lambda,T)=\int_{0}^{\infty}\frac{dV}{V_0} Q_N(V,T)e^{- \lambda V}
\end{eqnarray}
where $Q_N(V,T)$ is the partition function in the canonical ensemble, $\lambda=P/k_B T=\beta P$ 
and $V_0$ is the volume dimension parameter. In our work we use $V_0=1$ as in \cite{hill1987statistical}. However, it is easy to show, that our approach is valid for another possible factors such as $V_0^{-1}=\left(\partial\ln Q_N/\partial V\right)_{N,T}$ \cite{koper1996length} and $V_0=1/\lambda$ \cite{sack1959pressure}.

The partition function of $N$ particles in canonical ensemble can be written in terms of Mayers cluster integrals $b_k$ \cite{hill1987statistical}:
\begin{eqnarray}
\label{eq:PF_CE}
Q_N(V,T)=\sum_{\{n_k\}}\prod_{k=1}^{N}\frac{(V b_k)^{n_k}}{n_k!}\delta(N-\sum_{k=1}^{N}kn_k),
\end{eqnarray}

\noindent where the summation in \eqref{eq:PF_CE} is carried out over the partitions $\{n_1,...,n_N\}$ corresponding to $N=\sum_{k=1}^{N}kn_k$. The number of all partitions of a natural number is an extremely fast increasing function. For example, in the case of $N=137$ there are 11097645016 ways to write the set $\{n_1,...,n_{137}\}$. For this reason, assuming arbitrary sequence of $\{b_k\}_{k=1}^N$, the direct calculations of \eqref{eq:PF_CE} is impossible. However, we demonstrate that rigorous combinatoric result can be obtained for \eqref{eq:PF_IIE_1}.

For the further simplicity it is convenient to perform the following identical transformation for integrand in \eqref{eq:PF_IIE_1} and then to execute the Laplace transform over the volume $V$:  
\begin{widetext}
\begin{eqnarray}
\label{eq:PF_IIE_2}
\Delta=-\dfrac{\partial }{\partial \lambda}\int_{0}^{\infty}dV \frac{Q_N(V,T)}{V}e^{- \lambda V}=-\dfrac{\partial }{\partial \lambda} \sum_{\{n_k:N=\sum_{k=1}^{N}kn_k\}}\dfrac{(n_1+...+n_N-1)!}{n_1!n_2!...n_N!}
\prod_{k=1}^{N}\Big(\frac{b_k}{\lambda}\Big)^{n_k}
\end{eqnarray}
\end{widetext}
Let us analyze expression \eqref{eq:PF_IIE_2} considering it as a polynomial of $1/\lambda$. The degree of this polynomial can be obtained as follows: the sum $\sum_k n_k$ has the minimal value corresponding to the case when all $N$ molecules are contained in the largest possible cluster of size $N$, then $n_{i<N}=0$, $n_N=1$; the maximal value of the sum is defined in opposite case when all clusters are monomers, then $n_1=N$, $n_{i>1}=0$. Thus expression \eqref{eq:PF_IIE_2} can be represented as:
\begin{eqnarray}
\label{eq:PF_IIE_3}
\Delta=-\dfrac{\partial }{\partial \lambda} \sum_{k=1}^N \frac{a_k}{\lambda^{k}}= \frac{1}{\lambda^{N+1}}\prod_{i=1}^{N-1}\left(1-\frac{\lambda}{\lambda_i}\right)
\end{eqnarray}
where $\{a_k\}_{k=1}^{N}$ are the coefficients depending on $b_k$. As one can see from \eqref{eq:PF_IIE_3} the partition function $\Delta(\lambda)$ has exactly $N-1$ non zero roots $\{\lambda_i\}_{i=1}^{N-1}$ in complex $\lambda$-plane. Also, if $\lambda= \re \lambda+i \, \im \lambda$ is a root then its complex conjugate $ \re \lambda-i \, \im \lambda$ is also a root.

The partition function \eqref{eq:PF_IIE_3} has direct analogy with representation of grand canonical ensemble one in terms of Yang-Lee zeros \cite{yang1952statistical}. Indeed, using thermodynamic properties of NPT-ensemble, the chemical potential $\mu$ can be obtained from:
\begin{eqnarray}
\label{eq:chempot_1}
N\beta \mu=-\log \Delta=(N+1)\log\lambda-\sum_{i=1}^{N-1}\log\left(1-\frac{\lambda}{\lambda_i}\right)
\end{eqnarray}

\noindent In the thermodynamic limit the summation in (\ref{eq:chempot_1}) can be substituted by the integral over a certain region $\Gamma$ containing all zeros $\{\lambda_i\}_{i=1}^{N-1}$ in the complex plane, with a local density $\eta(\lambda)$.
\begin{eqnarray}
\label{eq:chempot_2}
N\beta \mu=(N+1)\log\lambda-\int_{\Gamma}d\lambda'\eta(\lambda')\log\left(1-\frac{\lambda}{\lambda'}\right)
\end{eqnarray}
 
The roots $\{\lambda_i\}_{i=1}^{N-1}$ can be calculated from  \eqref{eq:PF_IIE_2} after the transformations, for this reason some combinatorial preliminaries can be useful. The Newton's identity allows to express the power sums in terms of the elementary symmetric polynomials:
\begin{widetext}
\begin{eqnarray}
\label{eq:Newton}
p_N=\sum_{\{n_k:N=\sum_{k=1}^{N}kn_k\}}(-1)^N\frac{N(n_1+n_2+...n_N-1)!}{n_1!n_2!...n_N!}\prod_{k=1}^{N}(-e_k)^{n_k}
\end{eqnarray}
\end{widetext}
 
\noindent where $p_N(x_1,...,x_N)$ is the power sum symmetric polynomial, $e_k(x_1,...,x_N)$ is the elementary symmetric polynomial
\begin{eqnarray}
p_k(x_1,...,x_N)=\sum_{i=1}^{N}x_i^k \nonumber \\ 
e_k(x_1,...,x_N)=\sum_{1\le j_1<j_2...<j_k\le N}x_{j_1}...x_{j_k}
\end{eqnarray}

\noindent To use expression (\ref{eq:Newton}) in the calculations of (\ref{eq:PF_IIE_3}) we rewrite the elementary symmetric polynomials in terms of cluster integrals $b_k$ using the following polynomial equation:
\begin{eqnarray}
\label{eq:Master_Polynomial}
\lambda \xi^N-\sum_{k=1}^{N}b_k\xi^{N-k}=0
\end{eqnarray}
From Vieta's formulas one can obtain the following expression:
\begin{eqnarray}
\label{eq:e_k&b_k}
e_k(\xi_1,...,\xi_N)=(-1)^{k+1}b_k/\lambda
\end{eqnarray}
where $\xi_k$, $k=1,...,N$  are the roots of polynomial equation \eqref{eq:Master_Polynomial}. Thus, using expressions \eqref{eq:Newton} and \eqref{eq:e_k&b_k} the partition function \eqref{eq:PF_IIE_2} can be rewritten as follows:
\begin{widetext}
\begin{eqnarray}
\Delta=-\dfrac{1}{N}\dfrac{\partial }{\partial \lambda}\sum_{\{n_k:N=\sum_{k=1}^{N}kn_k\}}(-1)^N\frac{N(n_1+...+n_N-1)!}{n_1!n_2!...n_N!}\prod_{k=1}^{N}\left[-e_k(\xi_1,...,\xi_N)\right]^{n_k}, 
\end{eqnarray}
\begin{eqnarray}
\label{eq:IIPF_4}
\Delta=-\dfrac{1}{N}\dfrac{\partial p_N(\xi_1,...,\xi_N) }{\partial \lambda}.
\end{eqnarray}
\end{widetext}
Accounting for that the roots of (\ref{eq:Master_Polynomial}) $\xi_n\neq0$, for $n=1,...,N$, it is more convenient to consider the new variables $z_n=1/\xi_n$. Using definition of the power sum symmetrical polynomials one can obtain the partition function \eqref{eq:IIPF_4} in the terms of the roots $\{z_n\}_{n=1}^N$:
\begin{eqnarray}
\label{eq:IIPF_5}
\Delta=\sum_{n=1}^{N}\dfrac{\partial z_n}{\partial \lambda}z_n^{-(N+1)}=\sum_{n=1}^{N}\alpha_n(\lambda)z_n^{-(N+1)}
\end{eqnarray} 
where the roots $z_n$ can be found from the following polynomial:
\begin{eqnarray}
\label{eq:Master_Polynomials_2}
\sum_{k=1}^{N}b_k z_n^k-\lambda=0, \;\; \textmd{for} \;\; n=1,...,N 
\end{eqnarray} 
then the coefficients $\alpha_n$ can be represented as implicit function of $\lambda$ using (\ref{eq:Master_Polynomials_2})
\begin{eqnarray}
\label{eq:alpha}
\alpha_n(\lambda)=\frac{1}{\sum_{k=1}^{N}k b_k z_n(\lambda)^{k-1}}
\end{eqnarray} Thus, expression (\ref{eq:IIPF_5}) with equation (\ref{eq:Master_Polynomials_2}) allows to calculate the partition function in NPT-ensemble for any $\lambda$ and for finite number of molecules $N$ (it is useful for application to small systems \cite{aslyamov2014complex}).  

Obtained partition function (\ref{eq:IIPF_5}) is similar to the ones of Potts and hard hexagons models in terms of transfer matrix \cite{baxter2016exactly}. For this reason, the approaches which were developed to study the zeros of these partition functions can be used in the case of many-body systems in the NPT-ensemble. The zeros of Potts model were studied using Beraha-Kahane-Weiss theorem (in the form which was proved by Alan Sokal in \cite{sokal2004chromatic}). Accounting for $\alpha_n(\lambda)\neq0$, here it is more convenient to consider equimodular curve as in work \cite{assis2014integrability} dedicated to hard hexagons model with various boundaries conditions. In accordance to this approach, in the thermodynamic limit the partition function will have zeros when two or more maximum terms in the series (\ref{eq:IIPF_5}) have equal moduli \cite{assis2014integrability}: 

\begin{eqnarray}
\label{eq:equimodular}
|\alpha_i z_i^{-(N+1)}|=|\alpha_j z_j^{-(N+1)}|\geq\{|\alpha_k z_n^{-(N+1)}|\}_{n=1}^N,
\end{eqnarray}
where $i$ and $j$ are the indexes corresponding to maximum terms. This situation can be found near the singularities of the coefficient $\alpha_n$ defining by the following polynomial:
\begin{eqnarray}
\label{eq:multiple}
\sum_{k=1}^{N}kb_kz^{k-1}=0
\end{eqnarray} 
At the same time equation (\ref{eq:multiple}) is the z-derivative of polynomial (\ref{eq:Master_Polynomials_2}) and automatically becomes true at the case of multiple roots of (\ref{eq:Master_Polynomials_2}). Hence, the condition (\ref{eq:equimodular}) is satisfied at sufficiently small vicinity of multiple roots $z_j$ (\ref{eq:Master_Polynomials_2}). All $N-1$ multiple roots $z_j$ can be found from equation (\ref{eq:multiple}) and do not depend on $\lambda$. As shown above (\ref{eq:PF_IIE_3}) the partition functions in the NPT-ensemple has exactly $N-1$ zeros in the complex $\lambda$-plane. Thus, in the thermodynamic limit the zeros in the complex plane are defined by the following expression:
\begin{widetext}
\begin{eqnarray}
\label{eq:zeros_condition}
\lambda_j=\sum_{k=1}^{N}b_kz_j^k, \;\;\; \textmd{where} \;\;\;\sum_{k=1}^{N}k b_kz_j^{k-1}=0,\;\; j=1,...,N-1.
\end{eqnarray}
\end{widetext}

\begin{figure*}[t]
	\includegraphics[width=8cm]{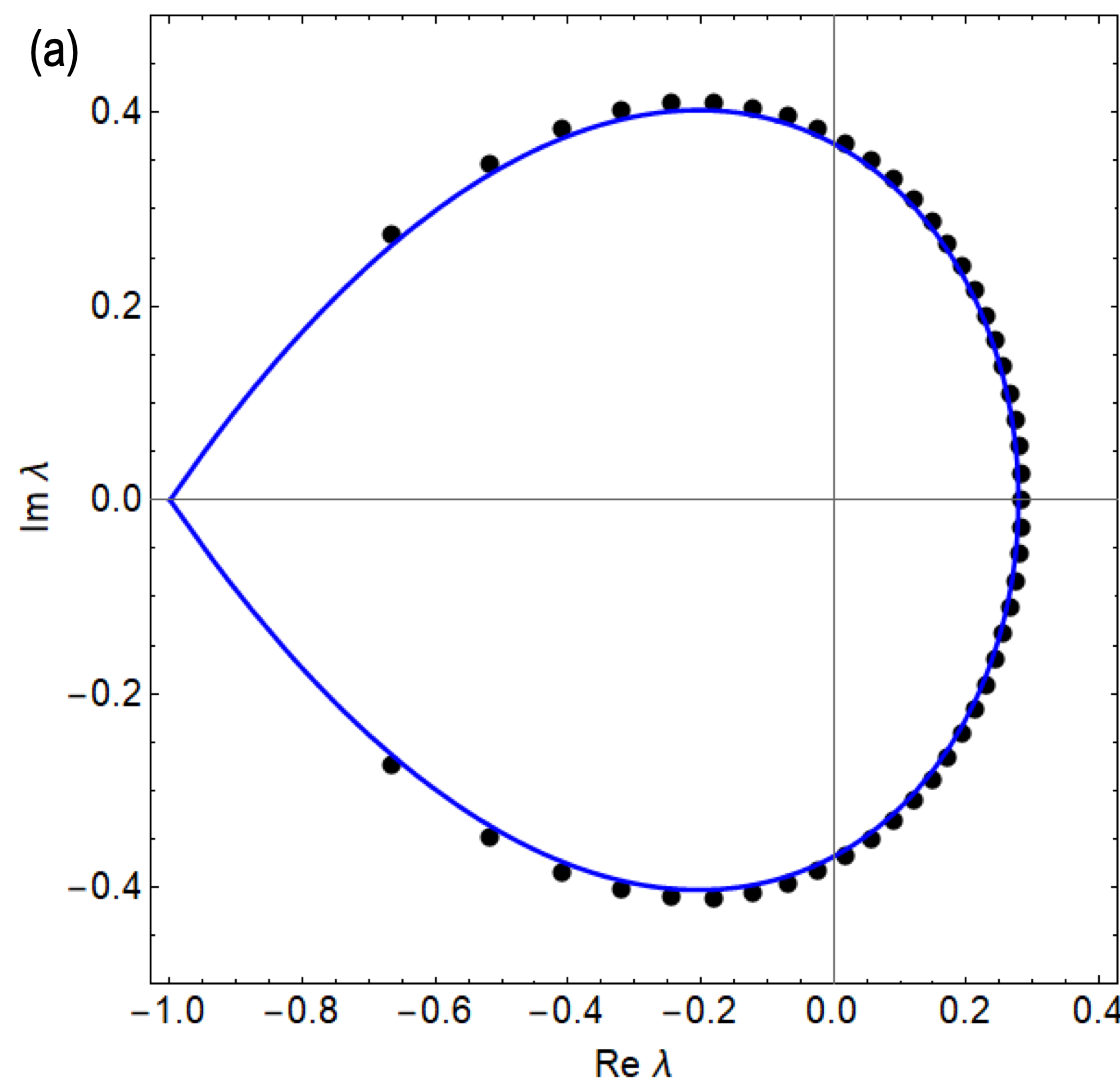}
	\includegraphics[width=8cm]{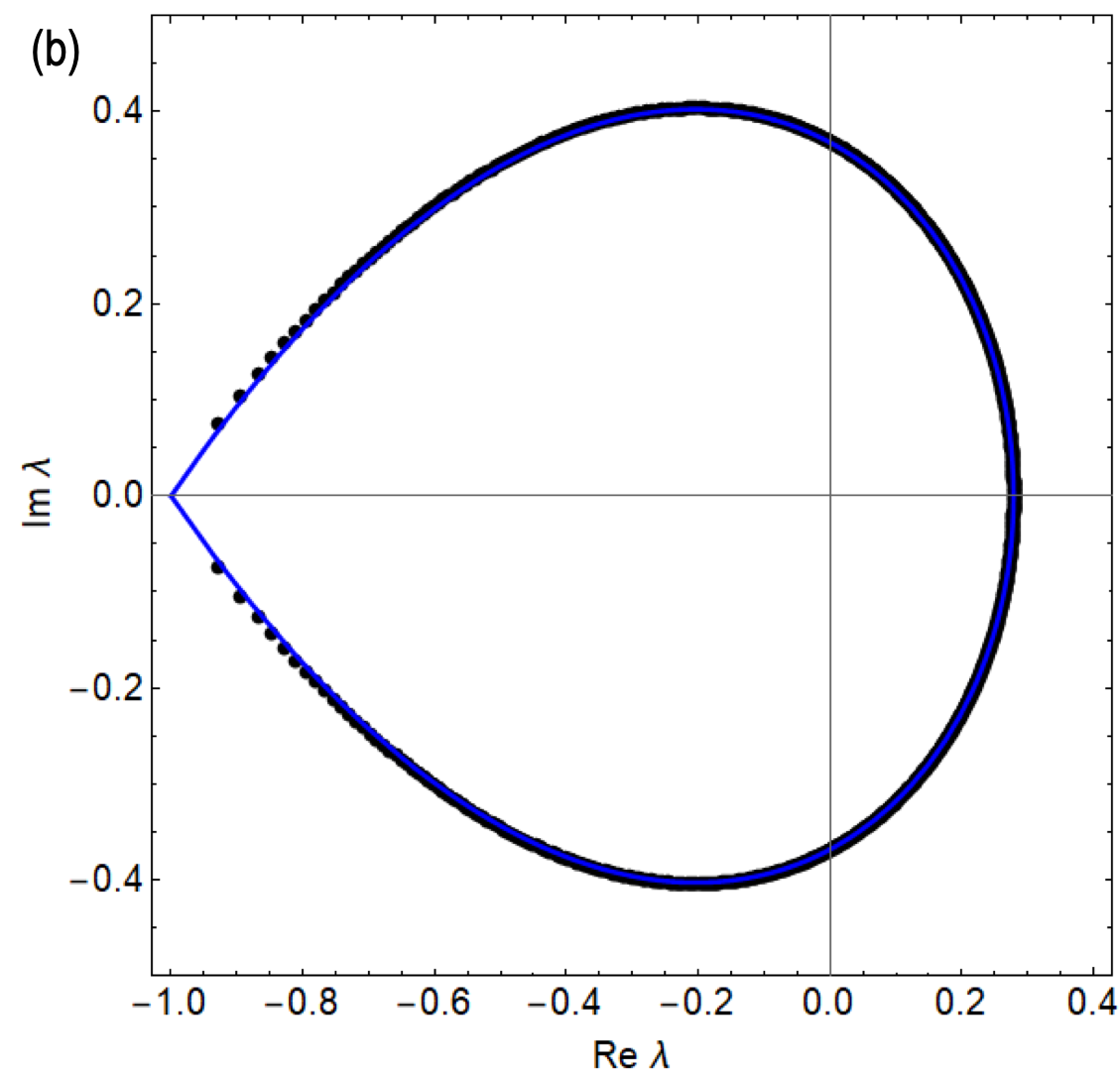}
	\caption{\label{fig:HS} Tonks gas zeros for various number of molecules and the limit curve. Black dots are zeros of (\ref{eq:HS_PF_1}) calculated from (\ref{eq:zeros_condition}) for the case of one dimensional hard sphere gas in the case of (a) $N=50$ and (b) $N=1000$. Blue curve is Szego curve (\ref{eq:Szego_Curve_Def}) after the substitution $\lambda\to-\zeta$.}
\end{figure*}

\section{Results}
To verify and illustrate the developed approach we considered the cases of one dimensional hard-sphere gas (Tonks gas) and for van der Waals (vdW) fluid. 

\subsection{Tonks gas}

In the case of hard-sphere gas, there is possible to describe NPT-ensemble by both our approach in terms of exactly known cluster integrals (\ref{eq:zeros_condition}) and by direct calculations of partition function using definition (\ref{eq:PF_IIE_1}). The exact solution of this system allows to formulate explicit equation of the limit curve on $\lambda$ plane for the zeros of the partition function in the thermodynamic limit.  For the system of one dimensional hard sphere gas the partition function in the canonical ensemble is exactly known:
\begin{eqnarray}
\label{HS_PF_CE}
Q^{HS}_N=\frac{(V-N)^N}{N!}.
\end{eqnarray}

\noindent Let us start with direct calculation of the NPT-ensemble partition function using definition (\ref{eq:PF_IIE_1}): 

\begin{eqnarray}
\label{eq:HS_PF_1}
&\Delta^{HS}=\frac{1}{N!}\int_{0}^{\infty}dVe^{-\lambda V}(V-N)^N= \\ \nonumber
&=\frac{1}{N!}e^{-\lambda N} \lambda ^{-N-1} \Gamma(N+1,-N \lambda)=\lambda ^{-N-1}e_N(-N\lambda)
\end{eqnarray}
\noindent where expressions $e_N(x)=\sum_{k=0}^{N}x^k/k!$ and $\Gamma(N+1,-N \lambda)=N! e^{N\lambda}e_N(-N\lambda)$ are used. Thus the zeros of the partition function (\ref{eq:HS_PF_1}) are defined by scaled exponential sum 
\begin{equation}
e_N(-N\lambda)=0    
\end{equation}

\noindent In 1924 Gabor Szego showed that the roots of $e_N(N \zeta)$ in the limit $N\to\infty$ trend to the following curve on complex $\zeta$-plane, $|\zeta|<1$ (now often titled as the Szego curve) \cite{pritsker1997szego}:
\begin{eqnarray}
\label{eq:Szego_Curve_Def}
|\zeta\exp(1-\zeta)|=1
\end{eqnarray} 
The zeros limit curve of (\ref{eq:HS_PF_1}) in the $\lambda$ plane can be easily obtained from (\ref{eq:Szego_Curve_Def}) after the following substitution $\lambda\to~-\zeta$, this curve is shown in Fig.~\ref{fig:HS}.

On the other hand, the exact expressions for cluster integrals for Tonks gas are well known \cite{caillol2003some}:
\begin{equation}
    b_n=\frac{(-n )^{n-1}}{n!}
\end{equation}
\noindent Then in accordance with our approach (\ref{eq:zeros_condition}) the zeros of the partition function (\ref{eq:HS_PF_1}) are defined as follows:
\begin{eqnarray}
\begin{cases}
\label{eq:zeros_HS}
\lambda_n^{(HS)}=\sum_{k=1}^{N}\frac{(-k )^{k-1}}{k!}z_n^k,
\\
\sum_{k=1}^{N}k \frac{(-k z_n)^{k-1}}{k!}=0, \;\; \text{where} \;\; n=1,...,N-1
\end{cases}
\end{eqnarray}
The results of calculation of $\{\lambda_n^{(HS)}\}_{n=1}^{N-1}$ for $N=50$ and $N=1000$ can be found in Fig.~\ref{fig:HS}. 
This figure demonstrates excellent agreement between developed approach (\ref{eq:zeros_HS}) and direct calculations of the partition function (\ref{eq:HS_PF_1}).

\subsection{van der Waals fluid}

One of the most popular model for real fluid is van der Waals equation of state, which can be written in the following form:
\begin{eqnarray}
\label{eq:vdW_EOS}
\lambda=\frac{\rho}{1-\rho}-\beta \rho^2
\end{eqnarray} 

\begin{figure}[t]
	\includegraphics[width=8cm]{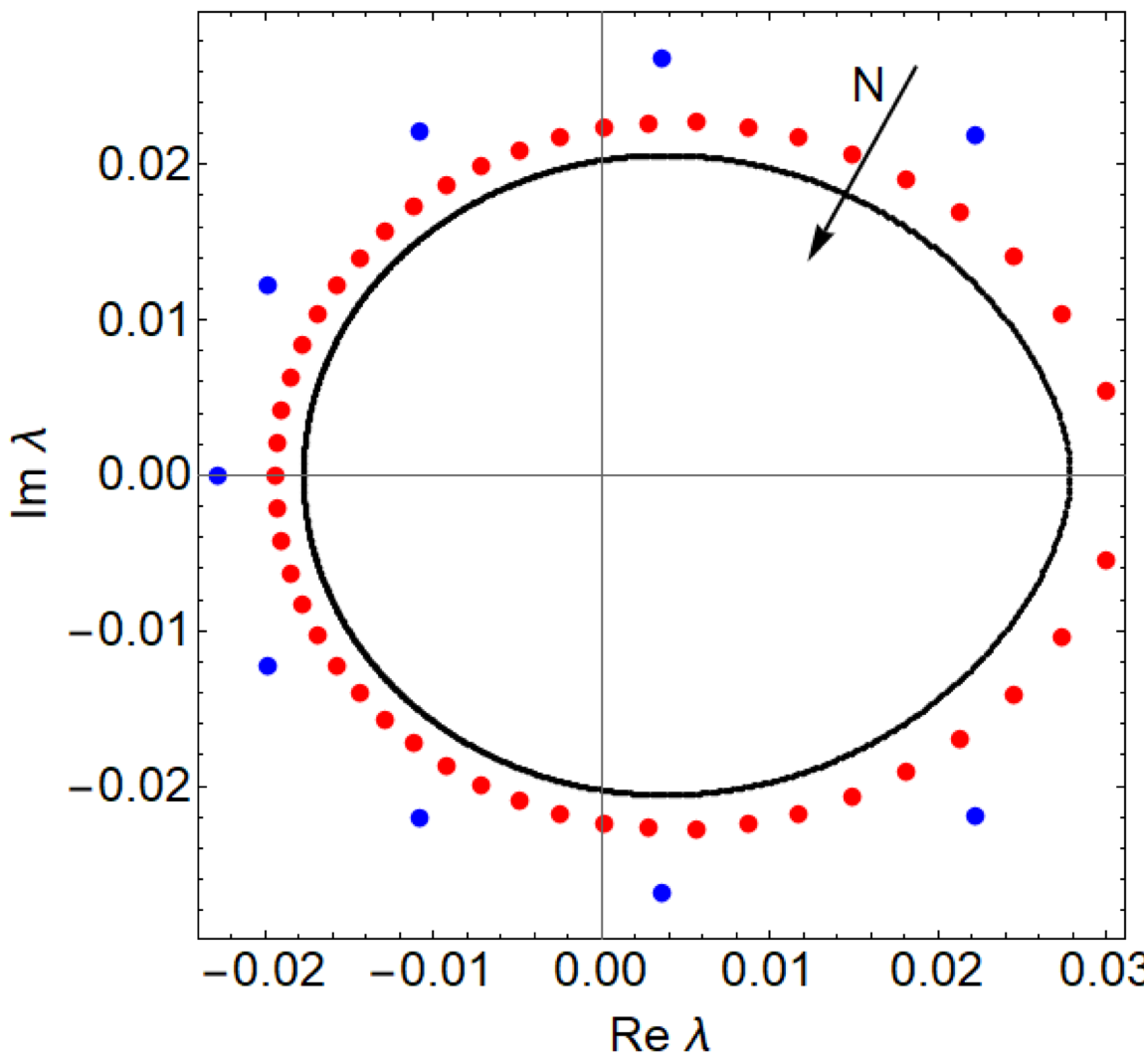}
	\caption{\label{fig:vdW_1} The zeros of vdW partition function in NPT ensemble at $\beta=10$ calculated for different number of molecules: blue, red and black correspond to $N=10$, $N=50$ and $N=1000$, respectively.}
\end{figure}

\begin{figure}[t]
	\includegraphics[width=8cm]{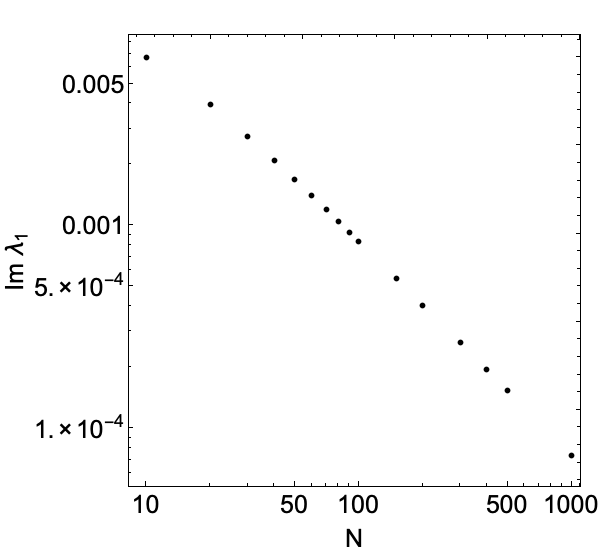}
	\caption{\label{fig:image_part} The complex zeros which satisfy the condition ($\re \lambda_1>\re \lambda_{i>1}$ and $\im \lambda_1>0$) are calculated for various $N$ at $\beta=10$.}
\end{figure}

It is well known that this model is able to describe phase transition for $\beta>\beta_c=27/8$. To consider this system in terms of zeros in the $\lambda$-plane the expressions for $b_n$ are needed. In the case of vdW fluids virial coefficients are known exactly: $B_1=1$, $B_2=1-\beta$, $B_{k>2}=1$ and the coefficients $b_n$ can be found from \cite{hill1987statistical}:
\begin{eqnarray}
\label{eq:bn}
b_n=\frac{1}{n^2}\sum_{\{n_k\}}\prod_{k=1}^{n-1}\frac{(n \beta_k)^{n_k}}{n_k!}\delta\left(\sum_{k=1}^{n-1}k n_k-n+1\right)
\end{eqnarray}
where $\beta_k=-B_{k+1}(k+1)/k $ are the irreducible cluster integrals. The first several coefficients can be calculated directly from (\ref{eq:bn}). In order to obtain $b_n$ for the large number of molecules we have accounted for in (\ref{eq:bn}) the most probable set $\{n_k^*\}$ only. Using Lagrange's method of undetermined multipliers, we have \cite{hill1987statistical}:
\begin{eqnarray}
\label{eq:Lagrange}
\frac{\partial }{\partial n_j}\sum_{k=1}^{n-1}\Big(n_k \log(n_k \beta_k)-n_k\log n_k+n_k+ \nonumber \\ 
+k n_k\log\alpha \Big)\Big\vert_{n_j=n_j^*}=0
\end{eqnarray}
where constant $\alpha$ can be found from the constraint $\sum_{k=1}^{n-1}k n_k=n-1$. Thus the most probable distribution is $n_k^*=n \beta_k\alpha^k$ and $\alpha$ is the smallest positive solution of the polynomial $(n-1)/n=\sum_{k=1}^{n-1}\beta_k \alpha^k$. Also, in work \cite{ushcats2017divergence} can be found an alternative approach to calculate (\ref{eq:bn}) using the recursive procedure.

\begin{figure}[t]
	\includegraphics[width=8cm]{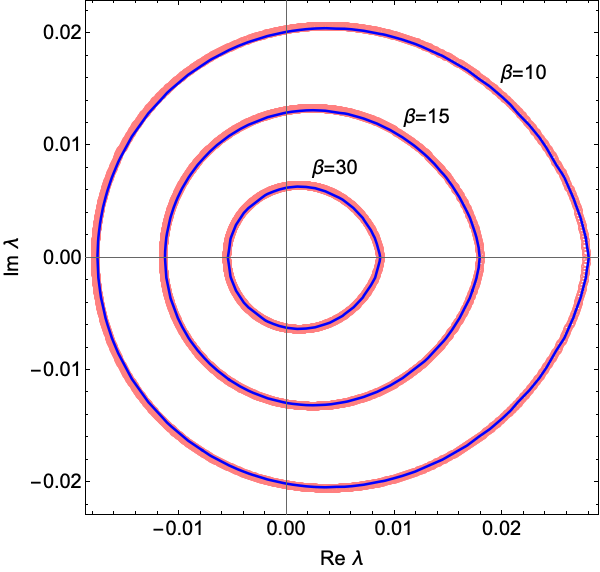}
	\caption{\label{fig:vdW_2} The comparison of combinatorial calculation of the zeros for $N=1000$ using (\ref{eq:zeros_condition}) (pink circles) and the limit curve in terms of modified Szego Curve (\ref{eq:LimitCurve}) (blue solid line)}
\end{figure}

Using above approximations for $b_n$ we studied vdW fluid in terms of the zeros in NPT-ensemble. Fig.~\ref{fig:vdW_1} and Fig.~\ref{fig:vdW_2} demonstrate representative examples of the zeros distributions at $\beta>\beta_c$. As one can see from Fig.~\ref{fig:vdW_1}, the zeros tend to the limit curve as the number of molecules $N$ increases. Thus at subcritical conditions the limit curve intersects the real axis $\lambda$ inducing a phase transition at certain pressure $P_\text{PT}$.  In Fig.~\ref{fig:image_part} the imaginary part of the zero which satisfies the condition ($\re \lambda_1>\re \lambda_{i>1}$ and $\im \lambda_1>0$) is shown as function of the number of molecules $N$. As one can see in the logarithmic coordinates the complex zero $\lambda_1$ almost linearly tends to the real axes. Thus, Fig.~\ref{fig:vdW_1} and Fig.~\ref{fig:image_part} result that the complex zeros strongly depends on $N$.

\begin{figure*}
	\includegraphics[width=17cm]{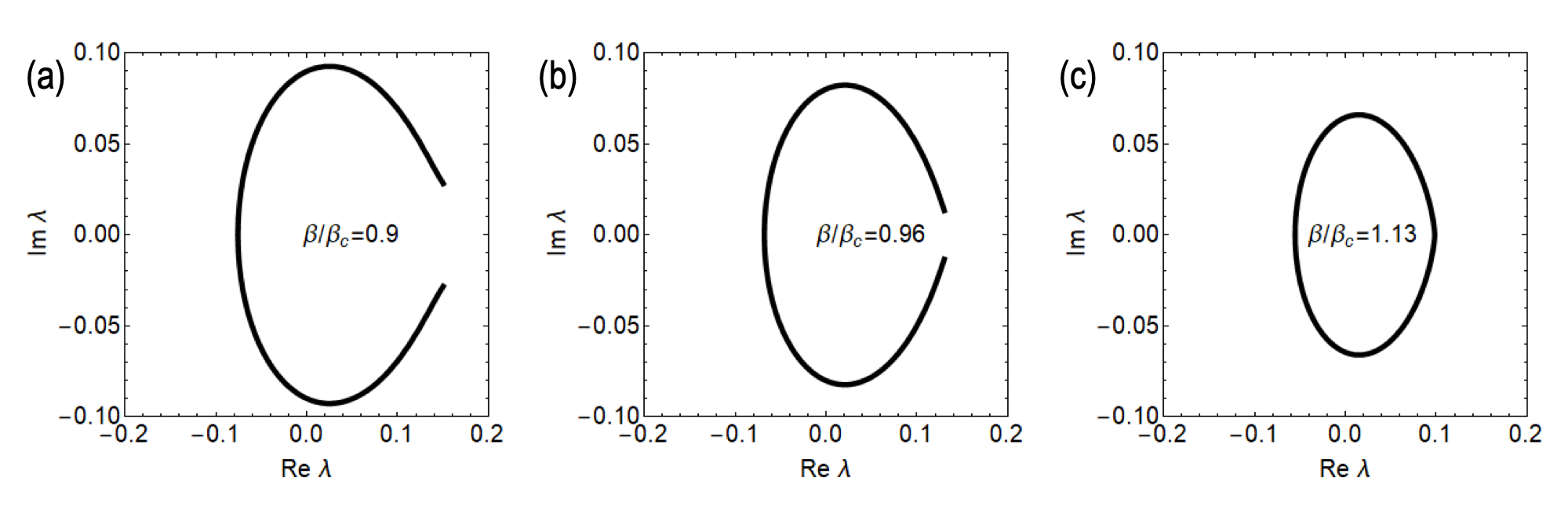}
	\caption{\label{fig:vdW_3} The limit curves for the zeros of vdW partition function in NPT-ensemble near critical temperature. When inverse temperature $\beta$ increases the limit curve tends to reach real the axis $\re \lambda$, from (a) to (c). At subcritical region (c) $\beta>\beta_c$ the curve intersects the real axis at certain phase transition point $\lambda_\text{PT}$}
\end{figure*}

To analyze the limit curve of vdW zeros in the $\lambda$-plane we applied Szego curve in modified form:
\begin{eqnarray}
\label{eq:SzegoCurve_Mod}
\big|f(\lambda)\exp\left[1-f(\lambda)\right]\big|=1
\end{eqnarray}
where $|f(\lambda)|\leq1$, the function $f(\lambda)$ is defined from fluid's EoS and equals to  $f(\lambda_\text{PT})=1$ at the point of phase transition. In case of vdW fluid the condition of phase transition can be represented using the compressibility as $\dfrac{\partial \lambda}{\partial \rho}=0$. Thus, possible modified Szego curve (\ref{eq:SzegoCurve_Mod}) is defined by the function $f(\lambda)=1-\dfrac{\partial \lambda}{\partial \rho}$ and has the following equation in terms of $\lambda$:
\begin{eqnarray}
\label{eq:LimitCurve}
\left|\left(1-\dfrac{\partial \lambda}{\partial \rho}\right)\exp\dfrac{\partial \lambda}{\partial \rho}\right|=1
\end{eqnarray}

Fig.~\ref{fig:vdW_2} shows the comparison of the zeros calculated by developed method (\ref{eq:zeros_condition}) for $N=10^3$ and the corresponding limit curve (\ref{eq:LimitCurve}) at different temperatures. As one can see from this figure, in all cases, analytical equation for the limit curve (\ref{eq:LimitCurve}) fits our combinatorial calculations well. Taking into account the lack of published data dedicated to the zeros in NPT-ensemble, the agreement of two independent theoretical approaches is a fair proof of the method adequacy.

Obtained equation (\ref{eq:LimitCurve}) for the limit curve can be used to characterize the nature of the phase transition which takes place when the zeros converge to some real point $\lambda_{PT}$. As was shown in (\ref{eq:PF_IIE_3})  the partition functions for NPT-ensemble in terms of $\lambda$ zeros has the same form as the result of Yang Lee for grand canonical ensemble \cite{lee1952statistical}. For this reason description of the phase transitions in terms of Yang Lee zeros can be applied to the zeros in the $\lambda$-plane \cite{bena2005statistical}. More precisely in case of vdW fluids there is the first order phase transition and the limit curve is a normal to the real axis at the crossing point. This fact can be checked using equation of the limit curve (\ref{eq:LimitCurve}) in a small vicinity of $\lambda_{PT}$. Indeed one can find that in terms of the density $\rho$ the limit curve (\ref{eq:LimitCurve}) in the vicinity of $\rho_{PT}$ is defined as $\rho=\rho_{PT}-\epsilon\pm i\epsilon$, where $\epsilon\to 0$. 
Then, using vdW equation (\ref{eq:vdW_EOS}), one can obtain the limit curve in vicinity of $\lambda_{PT}$ as:
\begin{eqnarray}
\label{eq:LimitCurve_PT}
&\re \lambda - \lambda_{PT}=O(\epsilon^3), \;\;\;\;\im \lambda=O(\epsilon^2)\nonumber \\
&\re \lambda - \lambda_{PT}\sim (\im \lambda)^{3/2}, \nonumber
\end{eqnarray}
where the tangent of the limit curve in $\lambda_{PT}$ is parallel to the imaginary axis. Fig.~\ref{fig:vdW_3} demonstrates the limit curve near critical point $\beta_c=27/8$. One can see that if $\beta/\beta_c<1$ there is not any limit point on positive real axis, and when $\beta/\beta_c>1$ the curve reaches a positive real value. 

\subsection{Conclusion}

Here we presented the theory of the zeros of the NPT-ensemble partition functions in the complex plane. Provided theory complements the powerful approaches Lee-Yang and Fisher zeros to describe the phase transitions in the most popular statistical ensembles. One of the most striking result of this work is an exact and an approximated equations for the zeros limit curves of Tonks gas and vdW fluid, respectively. Similarly with Lee-Yang circle theorem, these results revile fundamental properties of the complex thermodynamics. Since Lee-Yang zeros were already observed in experiments \cite{peng2015experimental, brandner2017experimental} our theory for NPT-ensemble extends the possible conditions for further experimental investigations of thermodynamics in the complex planes. Our approach is based on the cluster representation of the many-body partition functions which has often arisen in both classical and quantum systems. Therefore the developed theory can be also applied to the systems of the trapped Bose-Einstein gas \cite{mulken2001classification} and the nuclear droplet model \cite{moretto2005complement}. Actually the considered in this work case of van der Waals fluid can be extended to an arbitrary virial equation of state. Also, the concept of the partition function zeros is actively used in the study of nonequilibrium phase transitions \cite{heyl2018dynamical}. Therefore the information about NPT-ensemble can be useful in analysis of the Loschmidt amplitude zeros in the complex plane. For these reason the developed theory can be useful beyond an application to the considered molecular models.

\begin{acknowledgments}
	Authors are grateful to Konstantin Sinkov for discussions and useful comments. 
\end{acknowledgments}

\nocite{*}

\bibliography{sample}

\end{document}